\documentclass[aps,prl,reprint]{revtex4-1}
\usepackage{amsmath,amssymb} 
\usepackage{amsmath} 
\usepackage{amsfonts} 
\usepackage{bm} 
\usepackage[dvipdfmx]{graphicx} 
\usepackage{braket} 
\usepackage{physics}
\usepackage{color}

\begin{document}
\title{Emergent chiral symmetry in non-bipartite kagome and pyrochlore lattices with spin-orbit coupling}

\author{Hiroki Nakai$^1$}
\email{nakai-hiroki3510@g.ecc.u-tokyo.ac.jp}
\author{Masataka Kawano$^{1,2}$}
\author{Chisa Hotta$^1$}
\email{chisa@phys.c.u-tokyo.ac.jp}
\affiliation{$^1$Department of Basic Science, University of Tokyo, Meguro-ku, Tokyo 153-8902, Japan}
\affiliation{$^2$Department of Physics, Technical University of Munich, 85748 Garching, Germany}

\date{\today}

\begin{abstract} 
Chiral symmetry in energy bands appears as perfectly symmetric anti-bonding and bonding pairs of energy levels. 
It has only been observed in a few classes of models with a bipartite lattice structure or Bogoliubov-de-Gennes systems 
having the pairwise basis. 
We show that the {\it non-bipartite} kagome and pyrochlore lattices 
can host chiral symmetric bands when the strong spin-orbit coupling is introduced. 
There, the electrons hop to their neighbors by always converting the spin orientation up-side-down, 
which allows the up and down spin bases to form fictitious bipartite connections. 
The gauge invariant Wilson loop operator defined on a triangular unit serves as a marker to detect 
the presence of chiral symmetry, and using this property, the chiral operator is constructed. 
This allows us to access their topological symmetry classes that can easily change with small perturbations. 
\end{abstract}

\maketitle
{\it Introduction. } Possible types of electronic phases of matter are very often discussed and 
tabulated in terms of symmetries. 
The space group symmetries combined with time-reversal symmetry (TRS) 
determines the types of electronic band structures and the possible choices of symmetry broken phases out of them
\cite{Slater1972,Landau1999}. 
On the other hand, the periodic table developed for topological insulators (TI) and superconductors
has revealed that the symmetries that protect them and distinguish them from 
are not the spatial ones but are the TRS, particle-hole symmetry (PHS) and chiral symmetry (CS) \cite{Schnyder2008,Kitaev2009}. 
The CS is identified as the diphycercal shape of positive and negative parts in the energy spectrum 
and is present when the TRS and the PHS are both broken or both unbroken. 
These three symmetries generate a total of ten symmetry classes, which are extended 
through the combination with spatial symmetries \cite{Kruthoff2017}. 
\par
The CS serves as a guide to elucidate the nature of topological phases \cite{Wen1989}. 
For example, the winding number in the Su-Schrieffer-Heeger model \cite{SSH1979, SSH1980, Ryu2002} 
is defined using the CS operator, 
and the number of singular zero-mode Landau level and 
the bulk-edge correspondence of graphene \cite{Hatsugai2007, Hatsugai2009, Aoki2013} 
are understood by this symmetry. 
It explains the stable zero modes of bilayer graphene \cite{McCann2006,Katsnelson2008}, 
and when combined with spatial symmetry, it further protects the extra topological zero modes, 
showing us how many independent topological invariants are required to describe that state \cite{Koshino2014}. 
\par
Despite its importance, the CS is observed only in restricted classes of models; 
they are the bipartite hopping model \cite{Gade1991} and its analogues \cite{Kawarabayashi2011,Li2018}, 
the quadratic low energy effective Hamiltonian obeying the Bogoliubov-de-Gennes (BdG) equations 
with TRS and conserved magnetization, and the QCD models \cite{Verbaarschot1994}. 
How this symmetry could emerge in wider classes of physical systems is an important question. 
Here, we discover that the noninteracting fermions on a {\it non-bipartite} lattice with 
strong spin-orbit coupling (SOC) can host a CS band structure. 
We further prove that the gauge invariant Wilson loop operator \cite{Wen2004} serves as a detector of the CS, 
and using the Wilson loop operator, obtain an explicit form of the chiral operator. 
We show that several realistic types of perturbation to the CS energy bands will easily 
transform it from class DIII to various other classes in the topological periodic table; 
CS can be preserved or gives rise to the strong TI with apparent edge states. 
\begin{figure*}[t]
\begin{center}
\includegraphics[width=18cm]{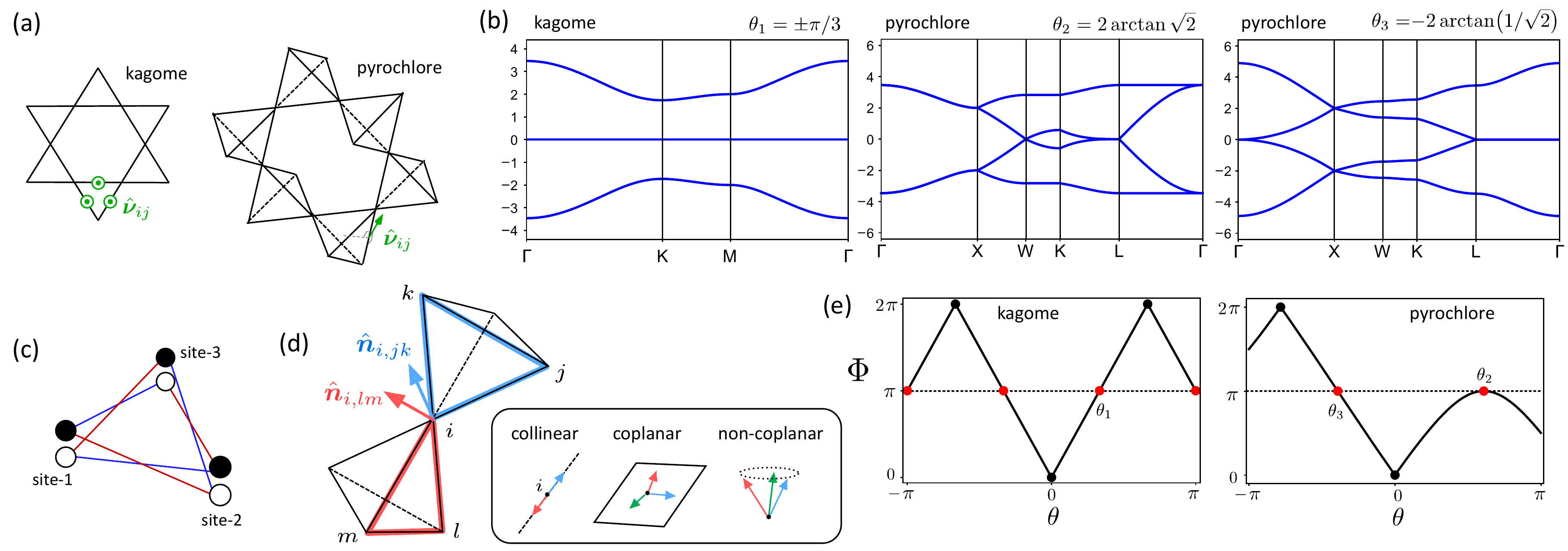}
\caption{(a) Kagome and pyrochlore lattices. 
The rotation axis $\hat{\vb*{\nu}}_{ij}$ of the SU(2) gauge field is shown. 
(b) The chiral symmetric band structures are shown at $\theta=\theta_1=\pm \pi/3$ and $\pm\pi$ for the kagome lattice 
and $\theta_2=2\arctan\sqrt{2}, \theta_3=-2\arctan(1/\sqrt{2})$ for the pyrochlore lattice, 
where $\lambda/t_0=\tan(\theta/2)$. 
(c) Schematic illustration of the fictitious bipartite structure on a triangle. 
The up/down spins are represented by black/white circles, 
respectively, and the different spins are connected. 
(d) The Wilson loop operator rotates the spins when hopping around the $C_{i,jk}$ loop 
by $\Phi$ about the $\hat{\vb*{n}}_{i,jk}$-axis. 
(e) $\Phi$ for the kagome and pyrochlore lattices.  
At CS points, $\theta_1,\theta_2,\theta_3$, we find $\Phi=\pi$.
}
\label{f1}
\end{center}
\end{figure*}
\par
{\it Chiral symmetry.} 
Without the loss of generality, the low energy effective Hamiltonian in momentum space is described by the 
quadratic Hamiltonian given as 
${\mathcal H}=  \sum_{\vb*{k}} \vb*{c}^{\dagger}_{\vb*{k}} \mathcal{H}(\vb*{k}) \vb*{c}_{\vb*{k}}$, 
where $\vb*{c}^{\dagger}_{\vb*{k}} 
= (c^{\dagger}_{\vb*{k}1\uparrow}, c^{\dagger}_{\vb*{k}1\downarrow}, \ldots,c^{\dagger}_{\vb*{k}n\uparrow}, c^{\dagger}_{\vb*{k}n\downarrow} )$ 
are the set of creation operators of a Bloch electron with spin $(\uparrow,\downarrow)$ 
and wavevector $\vb*{k}$ 
on the sublattice index $l=(1, ,2, \ldots, n)$, 
and the Bloch Hamiltonian $\mathcal{H}(\vb*{k})$ is the $2n\times2n$ Hermitian matrix. 
If the energy bands, $\pm E_m(\vb*{k})$, obtained by diagonalizing $\mathcal{H}(\vb*{k})$ 
is symmetric about the central zero energy level, 
there should exist a chiral operator $\Gamma$ that satisfies 
$\Gamma \mathcal{H}(\vb*{k}) \Gamma^\dagger = -\mathcal{H}(\vb*{k})$. 
In the basis that makes $\Gamma$ diagonal, 
the operators are represented as 
\begin{equation}
\Gamma = \mqty(I &  0 \\ 0 & -I ), \quad 
\mathcal{H}(\vb*{k}) = \mqty(0 &  D(\vb*{k}) \\ D^\dagger (\vb*{k}) & 0 ), 
\label{eq:chiralmat} 
\end{equation}
where $D(\vb*{k})$ is the $n_1\times n_2$ matrix in general, and there are at least $|n_1-n_2|$ zero modes. 
While for our Hamiltonian, we always find $n_1=n_2=n$. 
The eigenvectors of $\Gamma=\pm 1$ are described in the form, 
$(\vb*{\alpha}_m (\vb*{k}), \vb*{0})^T$ and $( \vb*{0}, \vb*{\beta}_m (\vb*{k}))^T$. 
Then, the energy eigenstates are their bonding or antibonding states given as 
\begin{equation}
\ket{\psi_{\pm m}(\vb*{k})} \propto  \mqty( \vb*{\alpha}_m (\vb*{k}) \\ \vb*{0} ) 
\pm \mqty( \vb*{0} \\ \vb*{\beta}_m (\vb*{k})), 
\label{eq:eivenv} 
\end{equation}
and $\Gamma$ interchanges them. 
The reason why we find a CS in the bipartite and BdG systems is that 
due to the equivalence of two sublattices or the superconducting pairs, 
the Bloch basis is trivially classified into two equivalent groups that have $\Gamma=\pm 1$. 
For our non-bipartite lattice, the up and down spins take this role, 
whereas remarkably, such spin-based CS does not necessarily require the spin-conversion-symmetry represented by TRS. 
\par
{\it SOC Hamiltonian on a non-bipartite lattice. } 
We consider a tight-binding Hamiltonian defined on the kagome and pyrochlore lattices given as 
\begin{equation}
\mathcal{H}= -t\sum_{\langle i,j \rangle} \vb*{c}^{\dagger}_{i} U_{ij} \vb*{c}_{j} + \mathrm{h.c.},
\label{eq:ham}
\end{equation}
where $\vb*{c}_{i}=(c_{i\uparrow}, c_{i\downarrow})^\mathrm{T}$ 
is the annihilation operator of an electron at site-$i$, 
and the summation is taken over the neighboring pairs of sites, $\langle i,j\rangle$. 
$\vb*{\sigma}=(\sigma_x, \sigma_y, \sigma_z)$ are Pauli matrices and 
the hopping amplitude is set to unity, $t=1$. 
When the electron hops from site-$j$ to the nearby site-$i$, 
its spin rotates by angle $\theta$ 
about the $\hat{\vb*{\nu}}_{ij}$-axis, which is expressed by the SU(2) gauge field 
$U_{ij}=\exp\left[-i\frac{\theta}{2}\hat{\vb*{\nu}}_{ij}\cdot\vb*{\sigma}\right]$, 
where $\hat{\vb*{\nu}}_{ji} = -\hat{\vb*{\nu}}_{ij}$ and $U_{ji} = U^\dagger_{ij}$. 
The unit vector $\hat{\vb*{\nu}}_{ij}$ is determined by the lattice symmetry, 
and points to the direction perpendicular to the plane for the kagome lattice \cite{Kim2015}, 
while points in different directions perpendicular to the bond $i$-$j$ 
for the pyrochlore lattice \cite{Kurita2011} (see Fig.\ref{f1}(a) and Supplementary A). 
The origin of the SU(2) gauge field is the SOC \cite{Shekhtman1992, Guarnaccia2012, Batista2014, Batista2020}; 
In many $4d$ and $5d$-electron systems, the interplay of strong {\it atomic} SOC and the crystal field 
enforces the reconstruction of energy levels on each site, 
and often the Kramers doublets labeled by the spin-index $\sigma$ form the valence bands
\cite{Kim2008, Uematsu2015, Khomskii2016, Takagi2021}. 
Examples in pyrochlore oxides that form such doublets are $d^5$ systems, $A_2$Ir$_2$O$_7$ \cite{Yanagishima2001, Matsuhira2007, Qi2012, Kondo2015} and Lu$_2$Rh$_2$O$_7$ \cite{Hallas2019}. 
In addition, $A_2$Os$_2$O$_7$\cite{Yamaura2012, Kataoka2022} and CsW$_2$O$_6$ \cite{Okamoto2020} 
of different valences also belong to this category 
in the presence of a trigonal crystal field \cite{Khomskii2016, Nakai2022}.
Since the spin momentum $\sigma$ is the combination of orbital angular momentum and electron-spin momentum, 
the electrons can hop between orbitals having different spins as, 
$i\lambda c_{i\alpha}^{\dagger} (\hat{\bm{\nu}}_{ij}\cdot\bm{\sigma})_{\alpha\beta}c_{j\beta}$.
By combining this spin-dependent hopping term with the ordinary hopping term, 
$-t_0 {c}_{i \alpha}^\dagger {c}_{j \alpha}$, the form Eq.(\ref{eq:ham}) 
is obtained as $t=\sqrt{t_0^2+\lambda^2}$ and 
$\theta=2\arctan(\lambda/t_0)$.  
\par
{\it SOC induced chiral symmetry. } 
As shown in Fig.\ref{f1} (b), 
the energy band structures of Eq.(\ref{eq:ham}) have CS 
at $\theta_1=\pm \pi/3,\pm\pi$ for the kagome lattice 
and $\theta_2=2\arctan\sqrt{2}, \theta_3=-2\arctan(1/\sqrt{2})$ for the pyrochlore lattice. 
To understand its origin, we need to find the form of two groups of basis sets 
that give $\Gamma=\pm 1$ in Eq.(\ref{eq:chiralmat}). 
Since both lattices consist of triangles, 
the basis set transformed back to the real space from the $\Gamma=\pm 1$ Bloch basis should be such that 
they have a finite hopping element between the neighboring sites only between different spin orientations. 
If such construction is possible, the up-spin electron on site-1 hops to site-2 and to site-3 by 
flipping its spin each time, and comes back to site-1 as a down-spin electron, 
as shown in Fig.\ref{f1} (c). If we go around the triangle twice, 
the initial spin orientation is recovered. 
This picture turns out to be valid when we construct the chiral operator. 
\par
{\it Gauge invariants. } 
We now define a parameter that detects the CS. 
To do so, we need to characterize the nature of the SU(2) gauge that this symmetry relies on. 
However, the explicit form of $U_{ij}$, namely $\theta$ and $\hat{\vb*{\nu}}_{ij}$ change 
if we apply the local gauge transformation: $\vb*{c}_{i} \rightarrow V_{i} \vb*{c}_{i}$, 
where $V_{i}$ denotes the SU(2) rotation of the quantized axis at site-$i$. 
This change is only a matter of representation and the band structure is gauge invariant. 
Accordingly, the parameter that characterizes the band structure should also be gauge invariant, 
for which we choose the Wilson loop operator: 
\begin{equation}
P(C_{i, jk}) = U_{ik} U_{kj} U_{ji} = \exp\left[ -i\frac{\Phi}{2} \hat{\vb*{n}}_{i,jk}\cdot\vb*{\sigma} \right],
\label{eq:wloop}
\end{equation}
constructed by the path-ordered product of SU(2) gauge field along the closed loop 
$C_{i,jk}: i \rightarrow j \rightarrow k \rightarrow i$ (see Fig.\ref{f1} (d)). 
When the electron circles along the loop $C_{i,jk}$, its spin orientation 
experiences a $\Phi$ rotation about the $\hat{\vb*{n}}_{i,jk}$-axis. 
Figure \ref{f1} (e) shows $\Phi$ as a function of $\theta$ 
for kagome and pyrochlore lattices. 
By comparing it with the band structures, 
we find that the CS is characterized by $\Phi=\pi$.  
For the kagome lattice, $\Phi$ and band structures are the same between $\pm \theta$, 
but not for the pyrochlore lattice. 
This is because $\hat{\vb*{\nu}}_{ij}$ is common to all bonds for the former and not for the latter, 
and accordingly, the $U_{ij}$ of different bonds are commutative/noncommutative for the former/latter, 
which we denote the Abelian/non-Abelian case. 
We comment that these two should be rigorously distinguished as such that there exists 
a gauge that makes the SU(2) gauge fields commutative in the Abelian case and no such gauge 
in the non-Abelian case. If the definition of non-Abelian depends on the choice of gauge \cite{Gao2020}, 
it does not give particular difference from the Abelian case about the gauge-invariant band structures 
or topological properties (see Supplementary B). 
When $\Phi=2\pi$, the flat-band is formed by the SOC in both lattices \cite{Nakai2022}. 
\par
For the non-Abelian case, there should be another gauge-invariant quantity that distinguishes the positive and negative $\theta$, 
which is the scalar product $\hat{\vb*{n}}_{i,jk}\cdot\hat{\vb*{n}}_{i,lm}$ 
around site-$i$ defined for different loops, $C_{i,jk}$ and $C_{i, lm}$. 
Although $\hat{\vb*{n}}_{i,jk}$ is not gauge invariant, their relative angle
is invariant (see Fig.~\ref{f1}(d) and proof in Supplementary B) \cite{Wen2004}. 
Using this fact, we can classify the states into four, 
which we call trivial, collinear, coplanar, and non-coplanar cases. 
For the trivial case, $\Phi=0, 2\pi$, the axis $\hat{\vb*{n}}_{i,jk}$ is not defined since 
$P(C_{i,jk}) \propto I$. 
For $\Phi \neq 0, 2\pi$, if all $\hat{\vb*{n}}_{i,jk}$ with different $jk$'s for the same $i$ 
are parallel we call it collinear, 
or if they are in the same plane we call it coplanar, and otherwise, it is non-coplanar
(see the inset of Fig.~\ref{f1}(d)). 
When the system is Abelian, it corresponds to either a collinear or trivial case. 
The non-Abelian case can be examined by comparing the two CS in Fig.~\ref{f1}(b): 
$\theta_2$ is collinear and $\theta_3$ is coplanar. 
\begin{figure}[tbp]
\begin{center}
\includegraphics[width=8.6cm]{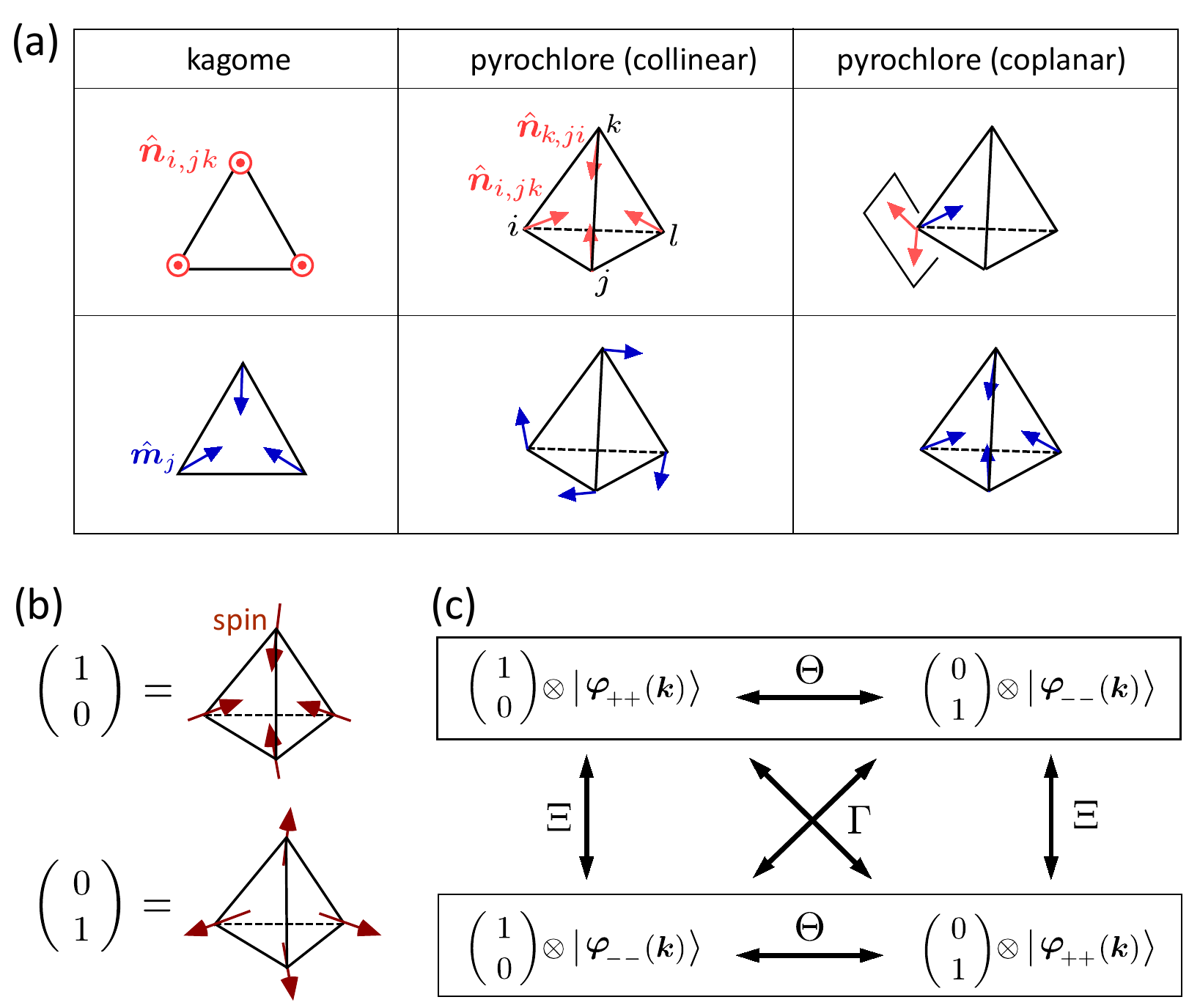}
\caption{(a) Vectors $\hat{\bm n}_{i,jk}$ and $\hat{\vb*m}_j$ on the kagome and pyrochlore lattices with CS. 
(b) All-in and all-out spin configurations forming a chiral pair. 
(c) Relationships of four different energy eigenstates of the pyrochlore bands (collinear type CS) 
that interchange by the TRS, PHS, and CS operations. }
\label{f2}
\end{center}
\end{figure}
\par
{\it Chiral operator. }
The general form of the local gauge transformation is 
$V_j=\exp\left[ -i\frac{\varphi_j}{2} \hat{\vb*{m}}_{j}\cdot\vb*{\sigma} \right]$, 
which rotates the spin quantization axis 
at site-$j$ by an angle $\varphi_j$ about the axis $\hat{\bm m}_j$. 
If one could find the particular form of $V_j$ that fulfills 
\begin{equation}
V_i U_{ij} V_j^\dagger= -U_{ij}, 
\label{eq:vuv}
\end{equation}
one can construct a chiral operator satisfying $\Gamma^2 =+I$ as 
\begin{equation}
\Gamma = \oplus_{j=1}^{n}  iV_j , 
\label{eq:gamma}
\end{equation}
for which we immediately find $\Gamma \mathcal{H}(\vb*{k}) \Gamma^\dagger = -\mathcal{H}(\vb*{k})$. 
To find such $V_j$ we focus on the fact that the Wilson loop operator in Eq.(\ref{eq:wloop}) 
is obtained by the product of three $U_{ij}$'s. Accordingly, the chiral operation 
that changes the sign of $U_{ij}$'s will change the sign of $P(C_{i, jk})$. 
Since $\Phi=\pi$ at the CS point, the sign-change is attained by the conversion of the axis 
via Eq.(\ref{eq:vuv}) as $\hat{\vb*{n}}_{i,jk}\rightarrow -\hat{\vb*{n}}_{i,jk}$. 
Namely, we need to set $\varphi_i=\pi$ and $\hat{\bm m}_i\perp \hat{\vb*{n}}_{i,jk}$ for all different $C_{i, jk}$'s around site-$i$. 
It follows that $\hat{\bm m}_i$ can be defined 
when and only when $\hat{\vb*{n}}_{i,jk}$'s on site-$i$ are either collinear or coplanar. 
To summarize, the chiral operator $\Gamma$ that satisfies Eq.(\ref{eq:chiralmat}) can be constructed 
when the Wilson loop operators satisfy: 
(i) $\Phi=\pi$, 
(ii) $\hat{\vb*{n}}_{i,jk}$ is collinear or coplanar, 
where we can write down the explicit form 
the form as $iV_j = \hat{\vb*{m}}_j \cdot \vb*{\sigma}$. 
\par
Figure~\ref{f2}(a) shows the actual directions of $\hat{\bm m}_j$ for three different CS. 
Then, Eq.~(\ref{eq:gamma}) indicates that 
in the basis that diagonalizes $\Gamma$, 
the up/down spin under the local quantization axis is parallel/antiparallel to $\hat{\vb*{m}}_j$. 
Because of the block-off-diagonal form of the Hamiltonian in Eq.(\ref{eq:chiralmat}), 
the electrons always turnover their spins when hopping. 
Namely, the up/down spin basis forms a fictitious bipartite connection 
which we discussed in Fig.~\ref{f1}(c). 
\par
{\it Emergent sublattice pseudospins. }
In describing the CS Hamiltonian, Eq.(\ref{eq:chiralmat}), 
we find $D^\dagger (\vb*{k}) \neq D(\vb*{k})$ in general, but 
our collinear case shows $D^\dagger (\vb*{k}) = D(\vb*{k})$ (Supplementary C). 
Then, for $\vb*{\varphi}_m(\vb*{k})$ that satisfies 
$D(\vb*{k}) \vb*{\varphi}_{m}(\vb*{k})=E_m  \vb*{\varphi}_{m}(\vb*{k})$, 
we find $\mathcal{H}(\vb*{k})\ket{\psi_{\pm m}(\vb*{k})} =\pm E_m\ket{\psi_{\pm m}(\vb*{k})}$, 
with $\ket{\psi_{\pm m}(\vb*{k})}$ in Eq.(\ref{eq:eivenv}) having 
$\vb*{\alpha}_m (\vb*{k})=\vb*{\beta}_m (\vb*{k})= \vb*{\varphi}_{m}(\vb*{k})$. 
\par
Reminding that the spin quantization axis for the CS basis 
is parallel to $\hat{\bm m}_j$ pointing toward the center of the triangle or tetrahedron, 
the eigenstates are the bonding and anti-bonding combinations of 
the all-in and all-out spin configurations shown in Fig.~\ref{f2}(b). 
In this construction, the sublattice(orbital) degrees of freedom in $D(\vb*{k})$ show some hidden symmetries;  
For the kagome lattice, 
\begin{equation}
D(\vb*{k})=\vb*{R}(\vb*{k}) \cdot \vb*{S},
\end{equation}
where $\vb*{S}$ is the spin-1 operator in the $3\times 3$ matrix representation 
(see Supplementary C). 
The eigenstate $\vb*{\varphi}_{m}(\vb*{k})$ of the sublattice degrees of freedom 
has three states with $S=1,0,-1$. 
The energy bands carry the sublattice pseudospins that point parallel/antiparallel to 
the SO(3) vector $\vb*{R}(\vb*{k})$ whose direction varies with $k$, 
which reminds us of a Rashba-Dresselhaus Hamiltonian \cite{Rashba1960,Dresselhaus1955}. 
\par
For the pyrochlore lattice, 
\begin{equation}
D(\vb*{k})=\vb*{R}_1(\vb*{k}) \cdot \vb*{S}_1 +\vb*{R}_2(\vb*{k}) \cdot \vb*{S}_2 ,
\label{eq:ds1s2}
\end{equation}
where $\vb*{R}_j(\vb*{k})$ is the three-dimensional vector. 
$\vb*{S}_1$ and $\vb*{S}_2$ are the spin-1/2 operators, which commute with each other 
in the $4\times 4$ matrix representation (Supplementary C). 
We find four eigenstates $\vb*{\varphi}_{m}(\vb*{k})$ depending 
on whether $\vb*{S}_1$ and $\vb*{S}_2$ are parallel or antiparallel to 
$\vb*{R}_1(\vb*{k})$ and $\vb*{R}_2(\vb*{k})$, 
which we denote as $\vb*{\varphi}_{\sigma_1 \sigma_2}(\vb*{k})\, (\sigma_i = \pm)$, 
and its eigenvalue is $(\sigma_1 |\vb*{R}_1(\vb*{k})| +\sigma_2 |\vb*{R}_2(\vb*{k})|)/2$.
\par
We briefly mention a few points about the chiral zero modes. 
The doubly degenerate chiral flat band of the kagome lattice appears 
because $n$ is odd \cite{Ramachandra2017}, 
differently from the Lieb lattice having $n_1\ne n_2$ in Eq.(\ref{eq:chiralmat}). 
For the pyrochlore lattice, the W and L point contacts for 
$\theta_2$ and the $\Gamma$-L nodal line for $\theta_3$ 
are the essential degeneracies \cite{Asano2011} protected by the CS and lattice symmetry 
(see Supplementary C using Eq.(\ref{eq:ds1s2})). 
\par
{\it Symmetries and perturbations. } 
How the TRS, PHS, and CS act on the energy eigenstates of the pyrochlore lattice 
at $\theta=\theta_2$ is shown in Fig.~\ref{f2}(c). 
$\Gamma$ flips the true spin but not the sublattice pseudospin $(\sigma_1,\sigma_2)$, 
and interchanges the energy level $\pm E_m$. 
$\Xi$ flips only the sublattice-spins and interchanges $\pm E_m$. 
$\Theta$ flips both and exchanges the energetically degenerate pairs of states. 
It then follows that if the system has the CS and TRS, it also has PHS, 
reproducing $\Xi=\Gamma \Theta$. 
Since $\Gamma$ is expressed as in Eq.(\ref{eq:gamma}), $\Theta^{-1} \Gamma \Theta = -\Gamma$, 
which indicates that $\Xi^2=+I$. 
Reminding that $\Xi$ takes the role of pseudo-TRS for sublattice pseudospins, 
$\Xi^2=+I$ prohibits the odd-numbered half-integer sublattice pseudospins. 
This explains why Eq.(\ref{eq:ds1s2}) is not described by a single pseudospin-$3/2$ 
but by the two pseudospin-$1/2$. 
\begin{table}[t]
	\caption{The effect of perturbations on TRS, PHS, and CS. 
The absence of symmetry is denoted as 0, 
and the presence of TRS or PHS is denoted as $\pm1$ according to $\Theta^2, \Xi^2=\pm I$, 
and that of CS as $1$. 
The last column shows the symmetry class the Hamiltonian belongs to \cite{Schnyder2008}.}
	\label{table:perturbation}
	\begin{center}
		\begin{tabular}{l rr cc}
			\hline\hline
		        perturbation  \rule{5mm}{0mm}& \rule{2mm}{0mm} TRS &\rule{2mm}{0mm} PHS & \rule{2mm}{0mm}CS \rule{2mm}{0mm}& \rule{2mm}{0mm}AZ class \\ \hline
			magnetic field & $0\;\;$ & $0\;\;$ & $0/1$ & A/AIII \\ 
			on-site potential & $-1\;\;$ & $0\;\;$ & $0$ & AII \\ 
			bond modulation & $-1\;\;$ & $+1\;\;$ & $1$ & DIII \\ 
			\hline\hline
		\end{tabular}
	\end{center}
\end{table}
\begin{figure}[tbp]
\begin{center}
\includegraphics[width=8.6cm]{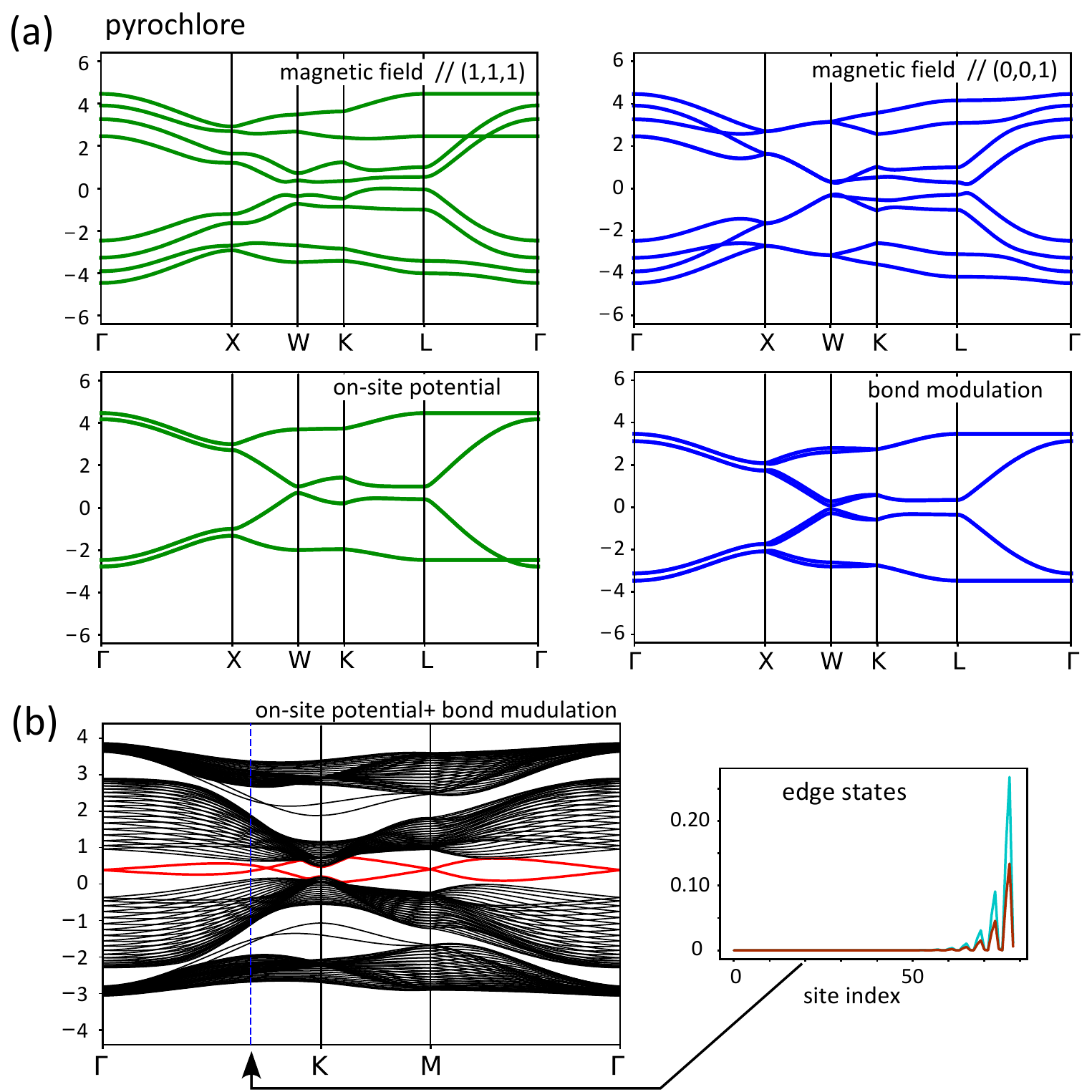}
\caption{(a) Band structures of the pyrochlore lattices when we add 
to Eq.(\ref{eq:ham}) the perturbations shown in Table I: 
magnetic field $h=1$ in two different directions, on-site potentials 
with $(w_1,w_2,w_3,w_4)=(1, 1, 1, 0.4)$, bond modulations. 
Green/blue bands are non-CS/CS. 
(b)Recalculated bands with on-site potentials and bond modulation, 
taking the inter-kagome direction an open boundary, 
where we find an edge state, indicated in red lines since this case corresponds to the strong TI in class AII. 
Distributions of edge modes (up/down spins for one mode and overwise for the other mode) in the open directions. }
\label{f3}
\end{center}
\end{figure}

\par
Another interest is how robust is our CS against the perturbations 
that may commonly appear in the actual material systems. 
Using the Wilson loop operator, 
we examine three types of perturbations as shown in Table \ref{table:perturbation} (see Supplementary D). 
The magnetic field (Zeeman terms) breaks both TRS and PHS, and for a particular 
choice of field direction, CS is preserved. 
This is in contrast to graphene, where the field generally preserves the CS \cite{Giesbers2007,Kawarabayashi2015}. 
The on-site potential keeps the TRS but breaks the PHS, and the CS is always broken. 
The bond modulation breaks the spatial inversion symmetry and the band degeneracy is lifted, 
but since it breaks neither TRS nor PHS, the CS is preserved. 
Several examples of band structures under these perturbations are shown in Fig.~\ref{f3}(a). 
We show in Fig.~\ref{f3}(b) the bands with on-site potentials and the bond modulation 
obtained by taking the inter-kagome plane direction an open boundary. 
The CS state belonged to class DIII transforms to class AII, where the strong TI is expected \cite{yongbaek2013}. 
Indeed we see a clear indication of the edge modes. 
\par
In conclusion, we found a new mechanism for generating 
a chiral symmetric band structure; 
the combination of SOC and the non-bipartite lattice structure based on the triangular loop units 
allows for the up-spin and down-spin subgroups of basis defined for properly chosen local quantization axis 
to form a fictitious bipartite lattice; 
when electrons hop to their neighbors they turnover their spin orientations 
and belong to the other group. 
Such condition for CS is detected using the gauge invariant Wilson loop operator on a triangle, 
and the chiral operator is defined as the product of local gauge transformation 
that flips the spin quantization axis upside-down. 
Although the magnetic field may seem to easily break the CS because 
the equivalence of up/down spin groups is lost, it does not.  
In our SOC-based CS, several different perturbations can transform the system to different 
classes in the topological table, which can be easily detected using the Wilson loop operator. 
Our CS can be extended from the doublet to the multiplet-based models, 
e.g. in the SU(4) model on a triangular lattice \cite{Yamada2021}. 

{\it Acknowledgment. }
This work was supported by a Grant-in-Aid 
for Transformative Research Areas "The Natural Laws of Extreme Universe---
A New Paradigm for Spacetime and Matter from Quantum Information" (No. 21H05191)
and JSPS KAKENHI (No.JP17K05533,JP21K03440).
M. K. was supported by JSPS Overseas Research Fellowship.
\bibliography{chiral}

\end{document}


\renewcommand{\figurename}{Figure S\!}

\title{Supplementary material for ``Emergent chiral symmetry in non-bipartite kagome and pyrochlore lattices with spin-orbit coupling"}
\author{Hiroki Nakai}
\affiliation{Department of Basic Science, University of Tokyo, Meguro-ku, Tokyo 153-8902, Japan}
\author{Masataka Kawano}
\affiliation{Department of Basic Science, University of Tokyo, Meguro-ku, Tokyo 153-8902, Japan}
\affiliation{Department of Physics, Technical University of Munich, 85748 Garching, Germany}
\author{Chisa Hotta}
\affiliation{Department of Basic Science, University of Tokyo, Meguro-ku, Tokyo 153-8902, Japan}

\maketitle
\begin{figure*}[tbp]
   \centering
   \includegraphics[width=18cm]{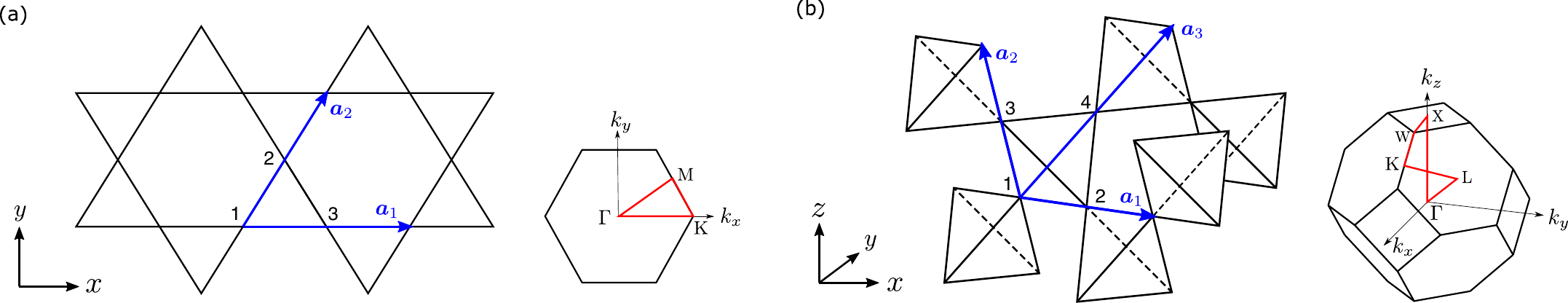}
  \caption{
   (a) Kagome lattice and 
   (b) pyrochlore lattice. 
   The corresponding Brillouin zone and the representative $k$-points are shown together. 
}
\vspace{-4mm}
\label{S1}
\end{figure*}
\subsection{Details of the SOC Hamiltonian}
We consider the SOC Hamiltonian in Eq.(3) in the main text, 
and show their explicit forms. 

\subsubsection{kagome lattice}
For the kagome lattice, 
we take the lattice vectors as 
$\vb*{a}_{1} = (1, 0)$, $\vb*{a}_{2} = \left(\frac{1}{2}, \frac{\sqrt{3}}{2} \right)$, 
and coordinates of three sublattices in the unit cell as 
$\vb*{r}_{1} = (0, 0)$, 
$\vb*{r}_{2} = \left(\frac{1}{4}, \frac{\sqrt{3}}{4} \right)$ and 
$\vb*{r}_{3} = \left(\frac{1}{2}, 0 \right)$ (Fig.S\ref{S1} (a)). 
In the corresponding first Brillouin zone in reciprocal space, 
the highly symmetric points are defined as 
$\mathrm{\Gamma} = (0, 0)$, 
$\mathrm{K} = \left( \frac{4\pi}{3}, 0 \right)$, and 
$\mathrm{M} = \left( \pi, \frac{\pi}{\sqrt{3}} \right)$. 
\par
The rotational axes of SU(2) gauge field in the Hamiltonian Eq.(3) 
for the kagome lattice using the above coordinates are given as, 
$\hat{\vb*{\nu}}_{12} = (0, 0, -1)$, 
$\hat{\vb*{\nu}}_{13} = (0, 0, 1)$, 
and 
$\hat{\vb*{\nu}}_{23} = (0, 0, -1)$. 
By constructing the basis in the order 
$( c^{\dagger}_{\vb*{k}1\uparrow}, c^{\dagger}_{\vb*{k}1\downarrow},
   c^{\dagger}_{\vb*{k}2\uparrow}, c^{\dagger}_{\vb*{k}2\downarrow},
   c^{\dagger}_{\vb*{k}3\uparrow}, c^{\dagger}_{\vb*{k}3\downarrow} )$, 
for ${\mathcal H}=  \sum_{\vb*{k}} \vb*{c}^{\dagger}_{\vb*{k}} \mathcal{H}(\vb*{k}) \vb*{c}_{\vb*{k}}$, 
the Bloch Hamiltonian is obtained in the $6\times 6$ form; 
\begin{equation}
\mathcal{H}(\vb*{k}) = \mqty(
0 & h_{12}(\vb*{k}) & h_{13}(\vb*{k})  \\
h_{21}(\vb*{k}) & 0 & h_{23}(\vb*{k})  \\
h_{31}(\vb*{k}) & h_{32}(\vb*{k}) & 0 
) ,
\end{equation}
where $h_{ij}(\vb*{k}) =-2tU_{ij}\cos(\vb*{k}\cdot(\vb*{r}_{i}-\vb*{r}_{j}))$.

\subsubsection{pyrochlore lattice}
We consider four sites in a unit cell with the lattice vectors 
given as
$\vb*{a}_{1} = \left(\frac{1}{2}, -\frac{1}{2}, 0 \right)$ , 
$\vb*{a}_{2} =  \left(0, -\frac{1}{2}, \frac{1}{2} \right)$, and 
$\vb*{a}_{3} =  \left(\frac{1}{2}, 0, \frac{1}{2} \right)$ (Fig.S\ref{S1} (b)). 
The corresponding reciprocal lattice vectors are
$\vb*{b}_{1} = 2\pi(1, -1, -1)$ , 
$\vb*{b}_{2} =  2\pi(-1, -1, 1)$, 
$\vb*{b}_{3} =  2\pi(1, 1, 1)$. 
The coordinates of four sublattices in the unit cell are given as
$\vb*{r}_{1} = \left(\frac{1}{8}, \frac{3}{8}, \frac{1}{8} \right)$ , 
$\vb*{r}_{2} = \left(\frac{3}{8}, \frac{1}{8}, \frac{1}{8} \right)$,  
$\vb*{r}_{3} = \left(\frac{1}{8}, \frac{1}{8}, \frac{3}{8} \right)$, and 
$\vb*{r}_{4} = \left(\frac{3}{8}, \frac{3}{8}, \frac{3}{8} \right)$. 
The highly symmetric points in the 1st Brillouin zone are 
$\mathrm{X} = (0, 0, 2\pi)$, 
$\mathrm{K} = \left( \frac{3\pi}{2}, 0, \frac{3\pi}{2} \right)$, 
$\mathrm{W} = ( \pi, 0, 2\pi )$, and $\mathrm{L} = (\pi, \pi, \pi)$. 
\par
The rotational axes of SU(2) gauge field are taken as 
$\hat{\vb*{\nu}}_{12} = (\frac{1}{\sqrt{2}}, \frac{1}{\sqrt{2}}, 0)$, $\hat{\vb*{\nu}}_{13} = (0, -\frac{1}{\sqrt{2}}, -\frac{1}{\sqrt{2}})$, 
$\hat{\vb*{\nu}}_{14} = (-\frac{1}{\sqrt{2}}, 0, \frac{1}{\sqrt{2}})$, $\hat{\vb*{\nu}}_{23} = (\frac{1}{\sqrt{2}}, 0, \frac{1}{\sqrt{2}})$, 
$\hat{\vb*{\nu}}_{24} = (0, \frac{1}{\sqrt{2}}, -\frac{1}{\sqrt{2}})$, $\hat{\vb*{\nu}}_{34} = (\frac{1}{\sqrt{2}}, -\frac{1}{\sqrt{2}}, 0)$,
and the representation of the Bloch Hamiltonian 
for the basis taken in the order, $(c^{\dagger}_{\vb*{k}1\uparrow}, c^{\dagger}_{\vb*{k}1\downarrow}, \ldots, c^{\dagger}_{\vb*{k}4\uparrow}, c^{\dagger}_{\vb*{k}4\downarrow} )$, is given as, 
\begin{equation}
\mathcal{H}(\vb*{k}) = \mqty(
0 & h_{12}(\vb*{k}) & h_{13}(\vb*{k}) & h_{14}(\vb*{k}) \\
h_{21}(\vb*{k}) & 0 & h_{23}(\vb*{k})  & h_{24}(\vb*{k})  \\
h_{31}(\vb*{k}) & h_{32}(\vb*{k}) & 0 & h_{34}(\vb*{k}) \\
h_{41}(\vb*{k}) & h_{42}(\vb*{k}) & h_{43}(\vb*{k}) & 0
) ,
\end{equation}
where $h_{ij}(\vb*{k}) =-2tU_{ij}\cos(\vb*{k}\cdot(\vb*{r}_{i}-\vb*{r}_{j}))$.

\subsection{Wilson loop operator}
The Wilson loop operator is the path-ordered product of SU(2) gauge fields $U_{ij}$ given in the main text as 
\begin{equation}
P(C_{i,jk}) = \exp\left[ -i\frac{\Phi}{2} \hat{\vb*{n}}_{i,jk}\cdot\vb*{\sigma} \right]. \hspace{10mm} (4)
\nonumber 
\end{equation}
In this section, we first derive the two gauge invariant properties from this definition. 
Then, we construct a chiral operator using the properties related to the Wilson loop operator. 
Finally, we clarify the definition of ``Abelian/non-Abelian'' used in this paper, 
which is different from the non-Abelian properties discussed in other works.  
We show an example that the Wilson loop operator can be used to determine the Abelian/non-Abelian properties.

\subsubsection{Proof of gauge invariance of $\Phi$ and $\hat{\vb*{n}}_{i,jk}\cdot \hat{\vb*{n}}_{i,lm}$}
We prove that the rotation angle $\Phi$ and the relative angle between rotation axes, 
$\hat{\vb*{n}}_{i,jk}\cdot \hat{\vb*{n}}_{i,lm}$, are both gauge invariant.
Using the local gauge transformation operator, 
$V_{i} = \exp\left[ -i\frac{\varphi_i}{2} \hat{\vb*{m}}_{i} \cdot\vb*{\sigma} \right]$, 
the Wilson loop operator is transformed as 
\begin{equation}
\begin{split}
V_{i} P(C_{i,jk}) V^{\dagger}_{i} &= \cos\frac{\Phi}{2} I -i(\hat{\vb*{n}}_{i,jk}\cdot\vb*{\sigma}) \cos\varphi_i \sin\frac{\Phi}{2} \\
&\quad -i(\hat{\vb*{m}}_{i}\cross\hat{\vb*{n}}_{i,jk}) \cdot\vb*{\sigma} \sin\varphi_i \sin\frac{\Phi}{2} \\
&\quad -i (\hat{\vb*{m}}_{i}\cdot\hat{\vb*{n}}_{i,jk})(\hat{\vb*{m}}_{i} \cdot\vb*{\sigma})(1-\cos\varphi_i)\sin\frac{\Phi}{2} \\
&= \cos\frac{\Phi}{2} I -i(\hat{\vb*{n}}'_{i,jk}\cdot\vb*{\sigma}) \sin\frac{\Phi}{2} \\
&= \exp\left[ -i\frac{\Phi}{2} \hat{\vb*{n}}'_{i,jk}\cdot\vb*{\sigma} \right], 
\end{split}
\end{equation}
where
\begin{equation}
\begin{split}
\hat{\vb*{n}}'_{i,jk} &= \cos\varphi_i \hat{\vb*{n}}_{i,jk} +\sin\varphi_i (\hat{\vb*{m}}_{i}\cross\hat{\vb*{n}}_{i,jk} ) \\
&\quad+(1-\cos\varphi_i) (\hat{\vb*{m}}_{i}\cdot\hat{\vb*{n}}_{i,jk}) \hat{\vb*{m}}_{i}.
\end{split}
\end{equation}
We immediately see that phase $\Phi$ remains unchanged. 
We now take the inner product between two rotation axes belonging to the same site-$i$ but to 
different loops, $C_{i,jk}: i \rightarrow j \rightarrow k \rightarrow i$ and $C_{i,lm}: i \rightarrow l \rightarrow m \rightarrow i$, as, 
\begin{equation}
\begin{split}
&\hat{\vb*{n}}'_{i,jk} \cdot \hat{\vb*{n}}'_{i,lm} \\
&= \cos^{2}\varphi_i \hat{\vb*{n}}_{i,jk}\cdot\hat{\vb*{n}}_{i,lm}
+\sin\varphi_i\cos\varphi_i (\hat{\vb*{m}}_{i}\cross\hat{\vb*{n}}_{i,jk} )\cdot\hat{\vb*{n}}_{i,lm}
\\
&\quad +\cos\varphi_i(1-\cos\varphi_i) (\hat{\vb*{m}}_{i}\cdot\hat{\vb*{n}}_{i,jk})(\hat{\vb*{m}}_{i}\cdot\hat{\vb*{n}}_{i,lm}) \\
&\quad +\sin\varphi_i\cos\varphi_i (\hat{\vb*{m}}_{i}\cross\hat{\vb*{n}}_{i,lm} )\cdot\hat{\vb*{n}}_{i,jk}
\\
&\quad +\sin^{2}\varphi_i (\hat{\vb*{m}}_{i}\cross\hat{\vb*{n}}_{i,jk} )\cdot (\hat{\vb*{m}}_{i}\cross\hat{\vb*{n}}_{i,lm} ) \\
&\quad +\sin\varphi_i(1-\cos\varphi_i)  (\hat{\vb*{m}}_{i}\cdot\hat{\vb*{n}}_{i,jk} ) \hat{\vb*{m}}_{i}\cdot (\hat{\vb*{m}}_{i}\cross\hat{\vb*{n}}_{i,lm} ) 
\\
&\quad +\cos\varphi_i(1-\cos\varphi_i) (\hat{\vb*{m}}_{i}\cdot\hat{\vb*{n}}_{i,jk})(\hat{\vb*{m}}_{i}\cdot\hat{\vb*{n}}_{i,lm}) \\
&\quad +\sin\varphi_i(1-\cos\varphi_i)  (\hat{\vb*{m}}_{i}\cdot\hat{\vb*{n}}_{i,lm} ) \hat{\vb*{m}}_{i}\cdot (\hat{\vb*{m}}_{i}\cross\hat{\vb*{n}}_{i,jk} )  
\\
&\quad +(1-\cos\varphi_i)^{2}(\hat{\vb*{m}}_{i}\cdot\hat{\vb*{n}}_{i,jk})(\hat{\vb*{m}}_{i}\cdot\hat{\vb*{n}}_{i,lm}).  
\end{split}
\end{equation}
The second and fourth terms cancel each other out, and the sixth and eighth terms are zero.
Putting together the third, seventh, and ninth terms, we get $(1-\cos^{2}\varphi_i)(\hat{\vb*{m}}_{i}\cdot\hat{\vb*{n}}_{i,jk})(\hat{\vb*{m}}_{i}\cdot\hat{\vb*{n}}_{i,lm})$.
Therefore, 
\begin{equation}
\begin{split}
\hat{\vb*{n}}'_{i,jk} \cdot \hat{\vb*{n}}'_{i,lm}
&= \cos^{2}\varphi_i \hat{\vb*{n}}_{i,jk}\cdot\hat{\vb*{n}}_{i,lm}
\\
&\quad+(1-\cos^{2}\varphi_i)(\hat{\vb*{m}}_{i}\cdot\hat{\vb*{n}}_{i,jk})(\hat{\vb*{m}}_{i}\cdot\hat{\vb*{n}}_{i,lm}) 
\\
&\quad +\sin^{2}\varphi_i(\hat{\vb*{m}}_{i}\cross\hat{\vb*{n}}_{i,jk} )\cdot (\hat{\vb*{m}}_{i}\cross\hat{\vb*{n}}_{i,lm} ),
\end{split}
\end{equation}
and
\begin{equation}
\begin{split}
&(\hat{\vb*{m}}_{i}\cross\hat{\vb*{n}}_{i,jk} )\cdot (\hat{\vb*{m}}_{i}\cross\hat{\vb*{n}}_{i,lm} ) 
\\
&= (\hat{\vb*{m}}_{i}\cdot\hat{\vb*{m}}_{i}) (\hat{\vb*{n}}_{i,jk} \cdot\hat{\vb*{n}}_{i,lm}) -(\hat{\vb*{m}}_{i}\cdot\hat{\vb*{n}}_{i,jk}) (\hat{\vb*{m}}_{i}\cdot \hat{\vb*{n}}_{i,lm}) \\
&=\hat{\vb*{n}}_{i,jk} \cdot\hat{\vb*{n}}_{i,lm}-(\hat{\vb*{m}}_{i}\cdot\hat{\vb*{n}}_{i,jk}) (\hat{\vb*{m}}_{i}\cdot \hat{\vb*{n}}_{i,lm}),
\end{split}
\end{equation}
and we obtain
\begin{equation}
\hat{\vb*{n}}'_{i,jk} \cdot \hat{\vb*{n}}'_{i,lm} =\hat{\vb*{n}}_{i,jk} \cdot\hat{\vb*{n}}_{i,lm}.
\end{equation}

\subsubsection{How to construct a chiral operator}
Let us explain how to construct the chiral operator when 
(i) $\Phi=\pi$ and (ii) $\{ \hat{\vb*{n}}_{i,jk} \}_i$ is collinear or coplanar. 
First of all, for rotation axes satisfying (ii), 
we can take a unit vector $\hat{\vb*{m}}_i$, which is perpendicular to all $\{ \hat{\vb*{n}}_{i,jk} \}_i$
that belong to site-$i$. 
Then, we find $V_i=\exp\left[ -i\frac{\pi}{2} \hat{\vb*{m}}_i \cdot \vb*{\sigma}\right] = -i\hat{\vb*{m}}_i \cdot \vb*{\sigma}$. 
Since $\hat{\vb*{m}}_i\cdot \hat{\vb*{n}}_{i,jk}=0$, 
\begin{equation}
\begin{split}
V_i P(C_{i,jk}) V^\dagger_i &= (-i\hat{\vb*{m}}_i \cdot \vb*{\sigma}) (-i\hat{\vb*{n}}_{i,jk} \cdot \vb*{\sigma}) (i\hat{\vb*{m}}_i \cdot \vb*{\sigma}) \\
&= + i\hat{\vb*{n}}_{i,jk} \cdot \vb*{\sigma} = -P(C_{i,jk}). 
\end{split}
\end{equation}
Next, to consistently generate the nearby $V_j\, (j\neq i)$ from $V_i$, 
the form 
\begin{equation}
V_j =U_{jk}U_{ki}V_i U_{ik}U_{kj}, \quad k\neq i,j,
\end{equation}
gives the natural extension using the SU(2) gauge. 
This form immediately gives $V_i U_{ij} V^\dagger_j =-U_{ij},\;\forall i,j$ e.g., 
\begin{equation}
\begin{split}
V_i U_{ij} V^\dagger_j &= V_i U_{ij} U_{jk} U_{ki} V^\dagger_i U_{ik} U_{kj} \\
&= V_i P(C_{i,kj}) V^\dagger_i U_{ik} U_{kj} \\
&= -P(C_{i,kj})U_{ik} U_{kj} \\
&= -U_{ij} U_{jk} U_{ki} U_{ik} U_{kj} \\
&= -U_{ij}, 
\rule{30mm}{0mm}
\end{split}
\end{equation}
and
\begin{equation}
\begin{split}
&\quad  V_l U_{lj} V^\dagger_j \\
&= U_{lm} U_{mi} V_i U_{im} U_{ml} U_{lj} U_{jk} U_{ki} V^\dagger_i U_{ik} U_{kj} \\
&= U_{lj} U_{ji} V_i P(C_{i, kj}) V^\dagger_i U_{ik} U_{kj} \\
&= -U_{lj} U_{ji} P(C_{i, kj}) U_{ik} U_{kj} \\
&= -U_{lj} U_{ji} U_{ij} U_{jk} U_{ki} U_{ik} U_{kj} \\
&= -U_{lj}. 
\end{split}
\end{equation}
In the second example, there appears a product of SU(2) gauge fields, $U_{im} U_{ml} U_{lj} U_{jk} U_{ki}$, 
along the path $i \rightarrow k \rightarrow j \rightarrow l \rightarrow m \rightarrow i$. 
Since the loops defined on the pyrochlore and triangular lattice cannot return 
to the original site by five independent bonds, 
two of the five bonds should be the same ones, which we chose as $j$-$l$ and $l$-$m$, i.e. 
$U_{ml} U_{lj}=1$. 
\par
We show another construction of the chiral operator, 
applied to a more general case including the parameter region off the chiral symmetric points. 
We take $\hat{\vb*{r}}_{Ci}$ as a unit vector that points from site-$i$ to the center of triangle or tetrahedron $C$. 
When $V_i=\exp\left[ -i\frac{\pi}{2} \hat{\vb*{r}}_{Ci} \cdot \vb*{\sigma}\right]$, 
\begin{equation}
V_i U_{ij}(\theta) V^\dagger_j = -U_{ij}(\alpha-\theta\pm 2\pi), 
\end{equation}
where $\alpha$ satisfy $\hat{\vb*{r}}_{Ci}\cdot\hat{\vb*{r}}_{Cj} =\cos(\alpha/2)$, $\hat{\vb*{r}}_{Ci}\cross\hat{\vb*{r}}_{Cj}=-\sin(\alpha/2) \hat{\vb*{\nu}}_{ij}$, and $0<\alpha<2\pi$. 
Because the unitary operator $\tilde{\Gamma}=\oplus_{j=1}^n =iV_j$ satisfy
\begin{equation}
\tilde{\Gamma} \mathcal{H}(\vb*{k}, \theta) \tilde{\Gamma}^\dagger = -\mathcal{H}(\vb*{k}, \alpha-\theta\pm 2\pi), 
\end{equation}
the energy eigenvalues of $\mathcal{H}(\vb*{k}, \theta)$ have opposite sign and the same amplitude 
as those of $\mathcal{H}(\vb*{k}, \alpha-\theta\pm 2\pi)$. 
Since the top bands are flat at $\theta=0$, 
flat bands appear at the bottom when $\theta=\alpha\pm2\pi$. 
If there exists $\theta_0$ that fulfills $\theta_0 =\alpha-\theta_0 \pm 2\pi$, 
$\mathcal{H}(\vb*{k}, \theta_0)$ has a chiral symmetry; 
it corresponds to $\theta_0=\arccos(-1/2)-\pi$ for kagome lattice, and $\theta_0=\arccos(-1/3)-\pi$ for pyrochlore lattice. 
In addition to that, since $\theta_0$ satisfies 
\begin{equation}
\Phi(\theta_0 +\varphi)=2\pi-\Phi(\theta_0 -\varphi)
\end{equation}
for arbitrary $\varphi$,
then $\Phi(\theta)$ is symmetric about $(\theta, \Phi)=(\theta_0, \pi)$. 


\subsubsection{Abelian and non-Abelian gauge fields}
We use the terminology, ``Abelian/non-Abelian" according to whether or not there exists a gauge that makes 
the SU(2) gauge fields $U_{ij}$ defined on different bonds commutative. 
This definition is more rigorous than the ones used in the previous articles; 
in general, when the SU(2) gauge field appears in the Hamiltonian, 
they simply call it ``non-Abelian"\cite{Gao2020}, 
but it can become Abelian with the aid of spatial symmetry, e.g. the site-centered inversion symmetry of a kagome lattice gives rise to the U(1) symmetry\cite{Kim2015}. 
However, even if $U_{ij}$ is ``non-Abelian", namely noncommutative {\it in some certain choices of the gauge}, 
if there is at least one way of taking the gauge 
that makes $U_{ij}$ commutative, 
the features linked to the noncommutativity do not appear in gauge-invariant physical quantities such as band structure or topological numbers. 
Since we are interested in the properties related to the non-commutativity of the gauge field, 
these properties can in turn enable us to judge whether the gauge is non-Abelian or not. 
\par
Let us show here that $\Phi$ is such property; 
If the gauge field is Abelian in our sense, 
one can simultaneously invert the rotation axis $\hat{\vb*{\nu}}_{ij}$ of all $U_{ij}$ 
by a local gauge transformation, which is equivalent to taking $\theta \rightarrow -\theta$. 
Then, since $\Phi$ is a gauge invariant quantity, 
one can say that if the gauge field is Abelian, $\Phi(\theta)=\Phi(-\theta)$ should hold. 
Conversely, if $\Phi(\theta) \neq \Phi(-\theta)$, the gauge field is non-Abelian. 
This condition works in Fig. 1(e) in the main text to find immediately that 
the SU(2) gauge field of the pyrochlore lattice is non-Abelian since $\Phi(\theta) \neq \Phi(-\theta)$. 


\subsection{Symmetry analysis}
In this section, we clarify several aspects of underlying symmetries of 
the energy eigenstates at chiral symmetric points.
First, we show that the U(1) symmetry in spin space allows us 
to block-diagonalize the Bloch Hamiltonian when it is collinear ($\theta=\theta_1$ in the kagome lattice 
and $\theta_2$ in the pyrochlore lattice), 
and that the diagonal block is described by a sublattice pseudospin due to chiral symmetry. 
Partially making use of the sublattice pseudospin picture, 
we next show that the chiral zero modes in the pyrochlore lattice 
are protected by the chiral symmetry and the spatial symmetry 
and appear at a specific point or line in momentum space, where the bands become degenerate.

\subsubsection{sublattice pseudospin}
Let us consider the collinear case, 
i.e., when the Wilson loop operators starting from site-$i$ represent the rotation around 
the same axis independent of their paths. 
Note that the rotation axis $\hat{\vb*{n}}_{i,jk}$ is different at the different base site $i$. 
Considering the local gauge transformation $V_i$ represented by the rotation around its axis, we find that $V=\oplus_{i=1}^n V_i$ is commutative with the Bloch Hamiltonian. 
It is confirmed in the same way as when we construct the chiral operator.
In the basis that diagonalizes $V$, the sublattices $i=1, 2, \ldots, n$ remain unchanged from the original basis. 
The Bloch Hamiltonian $\mathcal{H}(\vb*{k})$ is therefore block-diagonalized and expressed as $\mathrm{diag}( D(\vb*{k}), D^{*}(\vb*{k}))$ due to time reversal symmetry and site-centered inversion symmetry; 
$D(\vb*{k})$ is a $n\times n$ Hermitian matrix, 
and serves as a Hamiltonian about the sublattice degrees of freedom. 
We investigate the constraints on $D(\vb*{k})$ imposed by the chiral symmetry for the kagome lattice ($n=3$) and 
for the pyrochlore lattice ($n=4$).
\par
We first consider the kagome lattice, whose $D(\vb*{k})$ is expanded as
\begin{equation}
D(\vb*{k}) = \sum_{i=1}^{8} R_i(\vb*{k}) \tau_i,
\end{equation}
where $(\tau_1, \tau_2, \ldots, \tau_8)$ are Gell-Mann matrices:
\begin{equation}
\begin{split}
&\tau_1 = \mqty(0 & 1 & 0 \\ 1 & 0 & 0 \\ 0 & 0 & 0), \quad 
\tau_2 = \mqty(0 & -i & 0 \\ i & 0 & 0 \\ 0 & 0 & 0), \quad 
\tau_3 = \mqty(1 & 0 & 0 \\ 0 & -1 & 0 \\ 0 & 0 & 0), \\
&\tau_4 = \mqty(0 & 0 & 1 \\ 0 & 0 & 0 \\ 1 & 0 & 0), \quad 
\tau_5 = \mqty(0 & 0 & -i\\ 0 & 0 & 0 \\ i & 0 & 0), \quad 
\tau_6 = \mqty(0 & 0 & 0 \\ 0 & 0 & 1 \\ 0 & 1 & 0), \\
&\tau_7 = \mqty(0 & 0 & 0 \\ 0 & 0 & -i \\ 0 & i & 0), \quad 
\tau_8 = \frac{1}{\sqrt{3}}\mqty(1 & 0 & 0 \\ 0 & 1 & 0 \\ 0 & 0 & -2).
\end{split}
\end{equation}
Since the electron hops only between different sublattices, 
diagonal element of $\tilde{\mathcal{H}}(\vb*{k})$ is zero and $R_3(\vb*{k})=R_8(\vb*{k})=0$. 
\par
In the above given basis, 
the chiral operator $\Gamma$ is expressed as
\begin{equation}
\Gamma = \mqty( & \gamma \\ \gamma^{*} & ), \quad 
\gamma = {\rm diag}(z^{*}, i, -z), \quad z=e^{i\frac{\pi}{6}}.
\end{equation}
Since 
\begin{equation}
\Gamma \mathcal{H}(\vb*{k}) \Gamma^{\dagger} = \mqty(\dmat{ \gamma D(\vb*{k}) \gamma^{*}, \gamma^{*} D(\vb*{k}) \gamma }),
\end{equation}
the chiral symmetry, $\{ \mathcal{H}(\vb*{k}), \Gamma\}=0$, requires 
$\gamma^{*}D(\vb*{k}) \gamma = -D^{*}(\vb*{k})$.
Here, 
\begin{equation}
\begin{split}
\gamma^{*} \tau_1 \gamma &= -\frac{1}{2}\tau_1 -\frac{\sqrt{3}}{2} \tau_2, \quad \gamma^{*} \tau_2 \gamma = \frac{\sqrt{3}}{2}\tau_1 -\frac{1}{2} \tau_2, \\
\gamma^{*} \tau_4 \gamma &= -\frac{1}{2}\tau_4 +\frac{\sqrt{3}}{2} \tau_5, \quad \gamma^{*} \tau_5 \gamma = -\frac{\sqrt{3}}{2}\tau_4 -\frac{1}{2} \tau_5, \\
\gamma^{*} \tau_6 \gamma &= -\frac{1}{2}\tau_6 -\frac{\sqrt{3}}{2} \tau_7, \quad \gamma^{*} \tau_7 \gamma = \frac{\sqrt{3}}{2}\tau_6 -\frac{1}{2} \tau_7,
\end{split}
\end{equation}
and then,
\begin{equation}
\begin{split}
&\gamma^{*}D(\vb*{k}) \gamma \\
&= \big( -\frac{1}{2}R_{1}(\vb*{k}) +\frac{\sqrt{3}}{2}R_{2}(\vb*{k}) \big)\tau_1 
-\big(\frac{\sqrt{3}}{2}R_{1}(\vb*{k}) +\frac{1}{2}R_{2}(\vb*{k}) \big) \tau_2 \\
&\quad -\big(\frac{1}{2}R_{4}(\vb*{k})+\frac{\sqrt{3}}{2}R_{5}(\vb*{k}) \big)\tau_4 +\big( \frac{\sqrt{3}}{2}R_{4}(\vb*{k}) -\frac{1}{2}R_{5}(\vb*{k}) \big) \tau_5 \\
&\quad -\big( \frac{1}{2}R_{6}(\vb*{k}) -\frac{\sqrt{3}}{2}R_{7}(\vb*{k}) \big)\tau_6 
-\big( \frac{\sqrt{3}}{2}R_{6}(\vb*{k}) + \frac{1}{2}R_{7}(\vb*{k}) \big) \tau_7. \\
\end{split}
\end{equation}
On the other hand, since
\begin{equation}
\begin{split}
-D^{*}(\vb*{k}) &= -R_{1}(\vb*{k}) \tau_1 +R_{2}(\vb*{k}) \tau_2 -R_{4}(\vb*{k}) \tau_4 
+R_{5}(\vb*{k}) \tau_5 \\
&\quad -R_{6}(\vb*{k}) \tau_6 +R_{7}(\vb*{k}) \tau_7,
\end{split}
\end{equation}
we obtain three constraints
\begin{equation}
\begin{split}
&R_{1}(\vb*{k}) +\sqrt{3}R_{2}(\vb*{k}) = 0, \quad R_{4}(\vb*{k}) -\sqrt{3}R_{5}(\vb*{k}) = 0, \\
&R_{6}(\vb*{k}) +\sqrt{3}R_{7}(\vb*{k}) = 0.
\end{split}
\end{equation}
Using these constraints, we can reduce the number of 
basis to three, and by defining $\tilde{\vb*{R}}(\vb*{k})$ and $\vb*{S}$ as 
\begin{equation}
\begin{split}
\tilde{R}_1(\vb*{k}) &= \frac{\sqrt{3}}{2}R_{6}(\vb*{k}) -\frac{1}{2}R_{7}(\vb*{k}), \\
\tilde{R}_2(\vb*{k}) &= \frac{\sqrt{3}}{2}R_{4}(\vb*{k}) +\frac{1}{2}R_{5}(\vb*{k}), \\
\tilde{R}_3(\vb*{k}) &= \frac{\sqrt{3}}{2}R_{1}(\vb*{k}) -\frac{1}{2}R_{2}(\vb*{k}),
\end{split}
\end{equation}
\begin{equation}
\begin{split}
S_x &= \frac{\sqrt{3}}{2}\tau_6 -\frac{1}{2}\tau_7 = \mqty(0 & 0 & 0 \\ 0 & 0 & z \\ 0 & z^{*} & 0), \\
S_y &= \frac{\sqrt{3}}{2}\tau_4 +\frac{1}{2}\tau_5 = \mqty(0 & 0 & z^{*} \\ 0 & 0 & 0 \\ z & 0 & 0), \\
S_z &= \frac{\sqrt{3}}{2}\tau_1 -\frac{1}{2}\tau_2 = \mqty(0 & z & 0 \\ z^{*} & 0 & 0 \\ 0 & 0 & 0),
\label{eq:defs_kagome}
\end{split}
\end{equation}
$D(\vb*{k})$ is expressed as
\begin{equation}
D(\vb*{k}) = \sum_{i=1}^{3} \tilde{R}_i(\vb*{k}) S_i. 
\end{equation}
From Eq.(\ref{eq:defs_kagome}), we find that $\vb*{S}$ fulfills the 
commutation relation, 
\begin{equation}
[S_i, S_j] = i\varepsilon_{ijk}S_k, \quad \vb*{S}^{2} = 2I ,
\end{equation}
meaning that $\vb*{S}$ serves as a pseudospin-1 operator of the sublattice degrees of freedom.
\par
Next, we consider the pyrochlore lattice. 
In general, the Bloch Hamiltonian for pyrochlore lattice cannot be block-diagonalized because of 
its non-Abelian property. 
However, at $\theta=\theta_2=2\arctan(\sqrt{2})$, which is the collinear case, 
the system recovers the Abelian property and $\mathcal{H}(\vb*{k})$ can be block-diagonalized into
$\mathrm{diag}( D(\vb*{k}), D^{*}(\vb*{k}))$. 
Here, $D(\vb*{k})$ is expanded as
\begin{equation}
D(\vb*{k}) = \sum_{i=1}^{15} R_i(\vb*{k}) \Omega_i,
\end{equation}
where $(\Omega_1, \Omega_2, \ldots, \Omega_{15})$ are linearly independent traceless matrices:
\begin{equation}
\begin{split}
&\Omega_1 = \sigma_0\otimes\sigma_x, \quad 
\Omega_2 = \sigma_0\otimes\sigma_y, \quad 
\Omega_3 = \sigma_0\otimes\sigma_z, \\
&\Omega_4 = \sigma_x\otimes\sigma_x, \quad 
\Omega_5 = \sigma_x\otimes\sigma_y, \quad 
\Omega_6 = \sigma_x\otimes\sigma_z, \\
&\Omega_7 = \sigma_y\otimes\sigma_x, \quad 
\Omega_8 = \sigma_y\otimes\sigma_y, \quad 
\Omega_9 = \sigma_y\otimes\sigma_z, \\
&\Omega_{10} = \sigma_z\otimes\sigma_x, \quad 
\Omega_{11} = \sigma_z\otimes\sigma_y, \quad 
\Omega_{12} = \sigma_z\otimes\sigma_z, \\
&\Omega_{13} = \sigma_x\otimes\sigma_0, \quad 
\Omega_{14} = \sigma_y\otimes\sigma_0, \quad 
\Omega_{15} = \sigma_z\otimes\sigma_0. 
\end{split}
\end{equation}
As in the case of the kagome lattice, we set the coefficients as 
$R_7(\vb*{k})=R_{12}(\vb*{k})=R_{14}(\vb*{k})=0$, 
whose corresponding $\Omega_i$ have diagonal elements.
The chiral operator is expressed as 
\begin{equation}
\Gamma = \mqty( & \gamma \\ \gamma^{*} & ), \quad 
\gamma = {\rm diag}(z^{10}, z^{10}, z^{4}, z^{4}), \quad z=e^{i\frac{\pi}{6}}.
\end{equation}
The constraints arising from the chiral symmetry are derived in the same way as we did for the kagome lattice, 
and we find that
\begin{equation}
\gamma^{*} \Omega_i \gamma =
\begin{cases}  +\Omega_i & i=1, 2, 10, 11 \\ 
-\Omega_i & i= 4, 5, 6, 7, 8, 9, 13, 14,
\end{cases}
\end{equation}
and from $\gamma^{*}D(\vb*{k}) \gamma = -D^{*}(\vb*{k})$, we obtain 
\begin{equation}
\begin{split}
&R_1(\vb*{k})=R_5(\vb*{k})=R_7(\vb*{k})=R_9(\vb*{k})=R_{10}(\vb*{k})=R_{14}(\vb*{k})=0, 
\\
&\vb*{S}_1=(S_{1,x}, S_{1,y}, S_{1,z})=(-\Omega_2, -\Omega_6, \Omega_4)/2, \\
&\vb*{S}_2=(S_{2,x}, S_{2,y}, S_{2,z})=(-\Omega_{11}, -\Omega_{13}, \Omega_8)/2, 
\end{split}
\end{equation} 
where $\vb*{S}_1$ and $\vb*{S}_2$ are the pseudospin-$\frac{1}{2}$ operators, which satisfy, 
\begin{equation}
[S_{a,i}, S_{b,j}] = i\delta_{ab}\varepsilon_{ijk}S_{a,k}, \quad \vb*{S}_{a}^{2} = \frac{3}{4}I. 
\end{equation}

\begin{figure}[tbp]
	\centering
	\includegraphics[width=8.6cm]{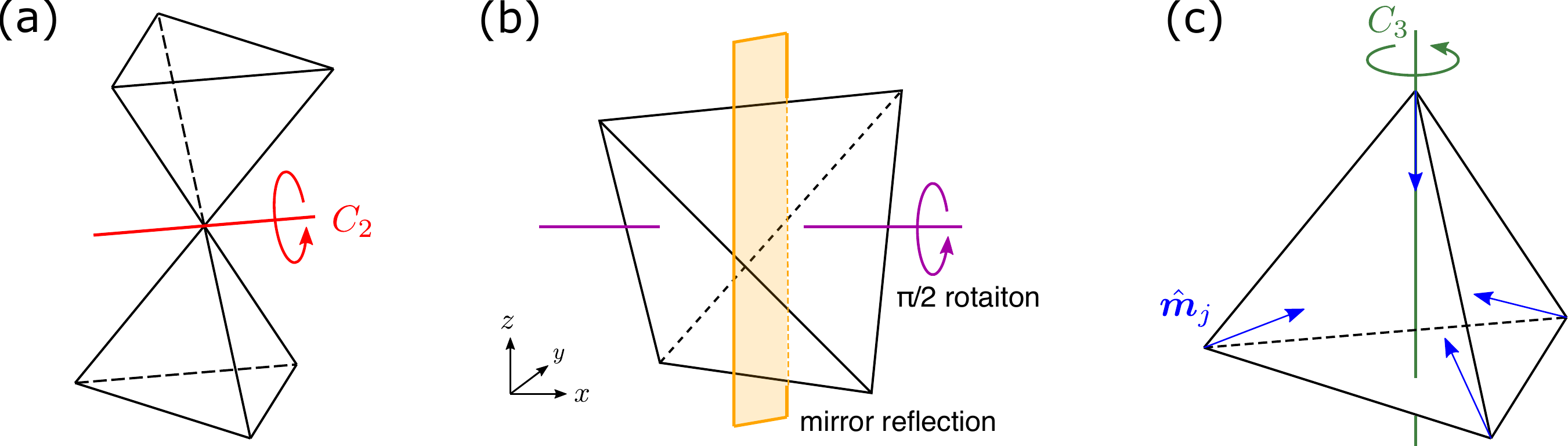}
	\caption{
		The symmetry operations. 
		(a) $C_2$ is a two-fold rotation, which interchange the upward and downward tetrahedra.
		(b) $s_{4\mu}\, (\mu=x,y,z)$ is a rotoreflection which is a combination of a $\pi/2$-rotation around the axis parallel to the $\mu$-axis through the center of the tetrahedron, and a mirror reflection across the plane perpendicular to that axis. 
		(c) $C_3$ is a three-fold rotation, which fix one vertex of the tetrahedron and replace the other three vertices. The quantization axes $\hat{\vb*{m}}_j$, which diagonalize the chiral operator, are invariant under this rotation.
	}
	\vspace{-4mm}
	\label{S2}
\end{figure}

\subsubsection{chiral zero modes}
We show the existence of the chiral zero modes protected by the chiral symmetry and the spatial symmetry. 
First, let us discuss the point contact at L and W points 
at $\theta=\theta_2=2\arctan(\sqrt{2})$. 
As discussed above, the Bloch Hamiltonian is block-diagonalized 
into $\mathrm{diag}( D(\vb*{k}), D^{*}(\vb*{k}))$ and is expanded as
\begin{equation}
\begin{split}
D(\vb*{k}) &= R_{2}(\vb*{k})\Omega_{2} +R_{4}(\vb*{k})\Omega_{4} +R_{6}(\vb*{k})\Omega_{6} \\
&\quad +R_{8}(\vb*{k})\Omega_{8} +R_{11}(\vb*{k})\Omega_{11} +R_{13}(\vb*{k})\Omega_{13}.
\label{eq:edeff_pyro}
\end{split}
\end{equation}
We now analyze the symmetry of the pyrochlore lattice. 
Let us set the axis perpendicular to the direction connecting one of the vertices and the center of the tetrahedron. 
This axis is parallel to one of the edges of the triangle formed by the rest of the vertices. 
The pyrochlore lattice is invariant under the $\pi$-rotations ($C_2$) about that axis (Fig.S\ref{S2} (a)).
The matrix representation of this operation for the basis that block-diagonalizes $\mathcal{H}(\vb*{k})$ is
\begin{equation}
\begin{split}
C_2\;\; &= \mqty( & \tilde{C}_{2, 1} \\ \tilde{C}_{2, 2} &), \\
\tilde{C}_{2, 1} &=\mqty(e^{\pi i/3} & & & \\ & & e^{5\pi i/6} & \\ & e^{5\pi i/6} & & \\ & & & e^{4\pi i/3} ), \\
\tilde{C}_{2, 2} &=\mqty(e^{2\pi i/3} & & & \\ & & e^{\pi i/6} & \\ & e^{\pi i/6} & & \\ & & & e^{5\pi i/3} ).
\end{split}
\end{equation}
Now, the L points in momentum space are invariant under $C_2$, 
and the Bloch Hamiltonian $\mathcal{H}(\vb*{k})$ commutes with $C_2$ as, 
\begin{equation}
\begin{split}
\mathcal{H}(\vb*{k}) &= C_2 \mathcal{H}(\vb*{k}) C^{\dagger}_2 \\
&= \mqty( & \tilde{C}_{2, 1} \\ \tilde{C}_{2, 2} &) \mqty(\dmat{D(\vb*{k}), D^{*}(\vb*{k})}) \mqty( & \tilde{C}^{\dagger}_{2, 2} \\ \tilde{C}^{\dagger}_{2, 1} &) \\
&= \mqty(\dmat{ \tilde{C}_{2, 1}D^{*}(\vb*{k}) \tilde{C}^{\dagger}_{2, 1}, \tilde{C}_{2, 2} D(\vb*{k}) \tilde{C}^{\dagger}_{2, 2} }).
\end{split}
\end{equation}
Then, we find 
\begin{equation}
D(\vb*{k}) = \tilde{C}_{2, 1}D^{*}(\vb*{k}) \tilde{C}^{\dagger}_{2, 1}, \quad D^{*}(\vb*{k}) = \tilde{C}_{2, 2} D(\vb*{k}) \tilde{C}^{\dagger}_{2, 2},
\end{equation}
which lead to
\begin{equation}
\begin{split}
& \tilde{C}_{2, 1} \Omega^{*}_{2} \tilde{C}^{\dagger}_{2, 1} = \Omega_{13}, \quad \tilde{C}_{2, 1} \Omega^{*}_{13} \tilde{C}^{\dagger}_{2, 1} = \Omega_{2}, \\
& \tilde{C}_{2, 1} \Omega^{*}_{4} \tilde{C}^{\dagger}_{2, 1} = \Omega_{8}, \quad \tilde{C}_{2, 1} \Omega^{*}_{8} \tilde{C}^{\dagger}_{2, 1} = \Omega_{4}, \\
& \tilde{C}_{2, 1} \Omega^{*}_{6} \tilde{C}^{\dagger}_{2, 1} = \Omega_{11}, \quad \tilde{C}_{2, 1} \Omega^{*}_{11} \tilde{C}^{\dagger}_{2, 1} = \Omega_{6}. 
\end{split}
\end{equation}
We thus find, $R_2(\vb*{k})=R_{13}(\vb*{k})$, $R_4(\vb*{k})=R_8(\vb*{k})$ and $R_6(\vb*{k})=R_{11}(\vb*{k})$. 
\par
The same analysis is performed for the other $C_2$ axes, and we obtain 
$R_2(\vb*{k})=-R_4(\vb*{k})=R_6(\vb*{k})=-R_8(\vb*{k})=R_{11}(\vb*{k})=R_{13}(\vb*{k}) (\equiv R(\vb*{k}))$. 
From these relationships, 
the block-diagonal Hamiltonian at L points is given as 
\begin{equation}
\begin{split}
D(\vb*{k}) &= R(\vb*{k}) (\Omega_2 -\Omega_4 +\Omega_6 -\Omega_8 +\Omega_{11} +\Omega_{13}) \\
&= -2 \vb*{R}(\vb*{k})\cdot(\vb*{S}_1 +\vb*{S}_2),
\end{split}
\end{equation}
where $\vb*{R}(\vb*{k})=R(\vb*{k})(1 ,1, 1)$, and its eigenvalues are $\pm 2|\vb*{R}(\vb*{k})|$, 0 (doubly degenerate). 
\par
The next symmetry operation we consider is the rotoreflection ($s_{4\mu}\, (\mu=x,y,z)$), the $\pi/2$-rotation about the axis parallel to the $\mu$-axis through the center of the tetrahedron
combined with the mirror operation 
about the plane perpendicular to that axis (Fig.S\ref{S2} (b)). 
Let us consider $\mu=x$ and this operation is represented by 
\begin{equation} 
\begin{split}
s_{4x} &= \mqty( & \tilde{s}_{4x, 1} \\ \tilde{s}_{4x, 2} &), \\
\tilde{s}_{4x, 1} &=\mqty( & e^{23\pi i/12} & & \\ &  & e^{17\pi i/12} & \\ &  & & e^{11\pi i/12} \\ e^{17\pi i/12} & & &  ), \\
\tilde{s}_{4x, 2} &=\mqty( & e^{19\pi i/12} & & \\ & & e^{\pi i/12} & \\ &  & &e^{7\pi i/12} \\e^{\pi i/12} & & & ).
\end{split}
\end{equation}
Now, the W points in momentum space are invariant under this operation and, similar to the above discussion at L points, we find
\begin{equation}
D(\vb*{k}) = \tilde{s}_{4x, 1}D^{*}(\vb*{k}) \tilde{s}^{\dagger}_{4x, 1}, \quad D^{*}(\vb*{k}) = \tilde{s}_{4x, 2} D(\vb*{k}) \tilde{s}^{\dagger}_{4x, 2},
\end{equation}
which lead to
\begin{equation}
\begin{split}
& \tilde{s}_{4x, 1} \Omega^{*}_{2} \tilde{s}^{\dagger}_{4x, 1} = -\Omega_{8}, \quad \tilde{s}_{4x, 1} \Omega^{*}_{8} \tilde{s}^{\dagger}_{4x, 1} = -\Omega_{2}, \\
& \tilde{s}_{4x, 1} \Omega^{*}_{4} \tilde{s}^{\dagger}_{4x, 1} = -\Omega_{11}, \quad \tilde{s}_{4x, 1} \Omega^{*}_{11} \tilde{s}^{\dagger}_{4x, 1} = \Omega_{4}, \\
& \tilde{s}_{4x, 1} \Omega^{*}_{6} \tilde{s}^{\dagger}_{4x, 1} = \Omega_{13}, \quad \tilde{s}_{4x, 1} \Omega^{*}_{13} \tilde{s}^{\dagger}_{4x, 1} = -\Omega_{6}. 
\end{split}
\end{equation}
We thus find, $R_{2}(\vb*{k}) = -R_{8}(\vb*{k}) (\equiv R(\vb*{k}))$, $R_{4}(\vb*{k}) = R_{6}(\vb*{k}) = R_{11}(\vb*{k}) = R_{13}(\vb*{k}) = 0$. 
From these relationships, 
the block-diagonal Hamiltonian at W points is given as 
\begin{equation}
D(\vb*{k}) = R(\vb*{k}) (\Omega_2 -\Omega_8) =-2R(\vb*{k}) ( S_{1,x} +S_{2,z}). 
\end{equation}
Since $[S_{1,x}, S_{2,z}]=0$, its eigenvalues are $\pm 2|R(\vb*{k})|$, 0 (doubly degenerate). 
\par
Finally, we consider the nodal $\mathrm{\Gamma}$-L line at $\theta=\theta_3=-2\arctan(1/\sqrt{2})$. 
Since it is the coplanar case, the block-diagonalization of ${\mathcal H}(\bm k)$ is unavailable, 
but by combining the chiral symmetry and the spatial symmetry, 
we can show that there is a band touching at zero energy \cite{Koshino2014}. 
Here, we consider the $2\pi/3$-rotation ($C_3$) about an axis that connects 
one vertex and the center of the tetrahedron. 
The matrix representation of this operation for the original basis is
\begin{equation}
C_3= \tilde{C}_3\otimes\mqty( & 1  &  &  \\ &  & 1  &  \\ 1 &  &  &  \\ &  &  & 1  ),
\end{equation}
where $\tilde{C}_3 = \exp\left[ -i\frac{\pi}{3} \hat{\vb*{n}} \cdot\vb*{\sigma} \right]$, $\hat{\vb*{n}}=(1, 1, 1)/\sqrt{3}$. 
In the basis that diagonalizes the chiral operator $\Gamma$, 
the quantization axis is $\hat{\vb*{m}}_j$ at site $j$. 
At $\theta=-2\arctan(1/\sqrt{2})$, it points toward the center of the tetrahedron. 
Therefore, this quantization axis is invariant under the $C_3$ operation, and $C_3$ commutes with $\Gamma$ (Fig.S\ref{S2} (c)). 
Since the $\mathrm{\Gamma}$-L line is invariant under $C_3$, 
$\mathcal{H}(\vb*{k})$ commutes with $C_3$. 
\par
Both $\Gamma$ and $\mathcal{H}(\vb*{k})$ that commute with $C_3$ 
are block-diagonalized into the eigenspace of $C_3$ with eigenvalues being 
$-1$, $\omega_1=e^{\pi i/3}$, $\omega_2=e^{5\pi i/3}$. 
They are, $\Gamma=\mathrm{diag}( \Gamma_{-1}, \Gamma_{\omega_1}, \Gamma_{\omega_2} )$ 
and $\mathcal{H}(\vb*{k})=\mathrm{diag}( \mathcal{H}_{-1}(\vb*{k}), \mathcal{H}_{\omega_1}(\vb*{k}), \mathcal{H}_{\omega_2}(\vb*{k}) )$, which are anticommutative in their respective blocks as 
\begin{equation}
\{ \Gamma_a, \mathcal{H}_a(\vb*{k})\} =0,\quad a=-1, \omega_1, \omega_2.
\end{equation}
$\nu_a = \Tr \Gamma_a$ denotes the difference in the number of zero modes, $N_{a,\pm}$, with positive and negative chirality (eigenvalues of $\Gamma_a$ is $\pm 1$); $\nu_a=N_{a,+} - N_{a,-}$. 
Since $\nu_{-1}=0$, $\nu_{\omega_1}=1$ and $\nu_{\omega_2}=-1$, there are at least two chiral zero modes. 
Since this argument holds for $\vb*{k}$-points on the $\mathrm{\Gamma}$-L line, we find that a nodal line appears on the $\mathrm{\Gamma}$-L line and that it is protected by the chiral symmetry and the three-fold rotation symmetry.

\subsection{Details of perturbation Hamiltonian}
We have discussed the effect of small perturbation to Eq.(3) in the main text, 
and summarized the results in Fig.3 and Table I. 
We explain the details of these perturbations. 
\subsubsection{uniform magnetic field}
We take account of the uniform magnetic field as a Zeeman effect as 
\begin{equation}
\mathcal{H}_{\rm mf} = -\sum_j \vb*{h}\cdot \vb*{s}_j, 
\end{equation}
where $s_{j, \mu}=\vb*{c}^\dagger_j (\sigma_\mu/2) \vb*{c}_j \, (\mu=x,y,z)$ is an electron spin operator 
at site-$j$. 
This magnetic field breaks both TRS and PHS, but may or may not break CS. 
We express the Zeeman term at site-$j$ as $-\vb*{h}\cdot\vb*{\sigma}=-i\exp\left[-i\frac{\pi}{2}\vb*{h}\cdot\vb*{\sigma}\right]$, and if $\vb*{h}$ is perpendicular to the quantization axis $\hat{\vb*{m}}_j$ diagonalizing the chiral operator, then $V_j (-\vb*{h}\cdot\vb*{\sigma}_j) V^\dagger_j =+\vb*{h}\cdot\vb*{\sigma} $ holds. 
When $\vb*{h}\cdot\hat{\vb*{m}}_j=0$ for each site,  TRS and PHS are violated, but CS is preserved.


\subsubsection{on-site potential}
We set different values of on-site potentials for $n$-independent sites in the unit cell 
\begin{equation}
\mathcal{H}_{\rm op} = \sum_{l=1}^n \sum_{j\in l} \omega_l n_j, 
\end{equation}
where $n_j=\vb*{c}^\dagger_j \vb*{c}_j$ is an electron number operator at site $j$. 
This on-site potential breaks the PHS but not the TRS, and thus the CS is lost.
This can also be interpreted as follows: 
the term at site-$j$ is expressed as $\omega_j I$, and then 
this term is shown to be invariant under the local gauge transformation as 
$V_j (\omega_j I) V^\dagger_j = \omega_j I \neq -\omega_j I$. 

\begin{figure}[tbp]
	\centering
	\includegraphics[width=5.6cm]{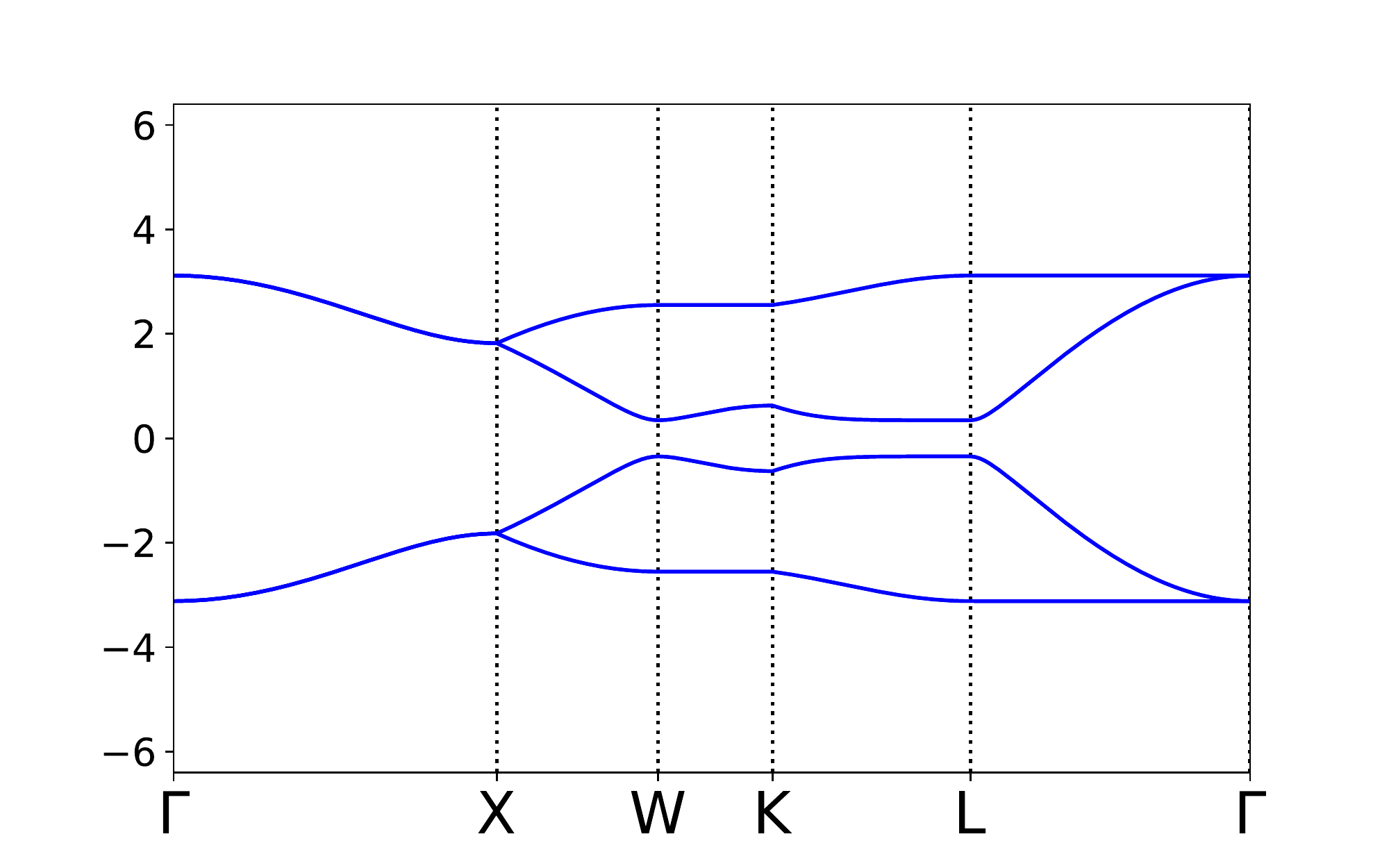}
	\caption{
		The band structure at $\theta_2$ for the pyrochlore lattice when bond modulation is $\delta_{ij}=1$ for all downward tetrahedra.
		The band degeneracy is not lifted even though the spatial inversion symmetry is broken.
	}
	\vspace{-4mm}
	\label{S3}
\end{figure}

\subsubsection{bond modulation}
The effect of bond modulation influences the hopping amplitude $t_0,\lambda$ in a nontrivial manner, 
because the crystal field is varied. 
However, for simplicity, we consider the simplest modulation called breathing: 
among the two types of tetrahedra pointing up and down, 
only one of them is subject to the modulation as 
$t^{\rm up}_{ij}=t$, $t^{\rm down}_{ij}=\delta_{ij}t$. 
This does not modify the SU(2) gauge field $U_{ij}$, which preserves the chiral symmetry. 
Moreover, since the TRS is not broken, the PHS is also preserved, 
and the class to which the Bloch Hamiltonian belongs remains unchanged. 
\par
This bond modulation breaks the spatial inversion symmetry of the lattice, 
which lifts the band-degeneracy in general. 
When $\delta_{12}=\delta_{34}$, $\delta_{13}=\delta_{24}$ and $\delta_{14}=\delta_{23}$, 
even though the space inversion symmetry is broken, 
it is not reflected in the band structure at $\theta_2$; the band degeneracy is not lifted (Fig.S\ref{S3}). 

\bibliography{supple}